# Ab Initio Molecular Dynamics on Quantum Computers

Dmitry A. Fedorov[1], Matthew J. Otten[2†], Stephen K. Gray[2], Yuri Alexeev[3]

[1]Oak Ridge Associated Universities, 100 Orau Way, Oak Ridge, TN 37830, USA. E-mail: dfedorov@nevada.unr.edu. ORCID: 0000-0003-1659-8580

[2]Center for Nanoscale Materials, Argonne National Laboratory, Lemont, IL 60439, USA.

[3]Computational Science Division, Argonne National Laboratory, Lemont, IL 60439, USA. E-mail: yuri@anl.gov, ORCID: 0000-0001-5066-2254

[†]Current address: HRL Laboratories LLC, Malibu, CA 90265, USA.

\***Corresponding Author contact information**: Yuri Alexeev; E-mail: yuri@anl.gov, Computational Science Division, Argonne National Laboratory, Lemont, IL 60439, USA.

## Abstract

Ab initio molecular dynamics (AIMD) is a valuable technique for studying molecules and materials at finite temperatures where the nuclei evolve on potential energy surfaces obtained from accurate electronic structure calculations. In this work, we present an approach to running AIMD simulations on noisy intermediate-scale quantum (NISQ) era quantum computers. The electronic energies are calculated on a quantum computer using the variational quantum eigensolver (VQE) method. Algorithms for computation of analytical gradients entirely on a quantum computer require quantum fault-tolerant hardware, which is beyond NISQ-era. Therefore, we compute the energy gradients numerically using finite differences, the Hellmann-Feynman theorem (HFT), and a correlated sampling technique. This method only requires additional classical calculations of electron integrals for each degree of freedom, without any additional computations on a quantum computer beyond the initial VQE run. As a proof of concept, AIMD dynamics simulations are demonstrated for the H$_2$ molecule on IBM quantum devices. In addition, we demonstrate the validity of the method for larger molecules using full configuration interaction (FCI) wave functions. As quantum hardware and noise mitigation techniques continue to improve, the method can be utilized for studying larger molecular systems.

## Introduction

The time evolution of molecular and material systems has been carried out for decades on classical computers using quantum and molecular dynamics (MD) calculations.[1–10] There are different levels of complexity and accuracy of such calculations. The most accurate and computationally demanding method is, of course, full quantum dynamics, where the time-dependent Schrödinger equation (TDSE) is solved for both electrons and nuclei. Often such approaches are formulated in terms of coupled Born-Oppenheimer potential energy surfaces for the electronic degrees of freedom, and various approximations are introduced to make the problem in question tractable.[11] Depending on the method, the system size that can be studied in quantum dynamics can vary from several degrees of freedom to 50 atoms. On the other end of the complexity spectrum is the field of classical MD methods, which is capable of describing chemical systems consisting of thousands of atoms.[2] One reason that the classical MD of such large systems is possible is that the movement of atoms is described classically through the solution of Hamilton's equations. Another reason is that the calculations of the interatomic forces required for solving Hamilton's equations are computed from empirical potential functions, also called force fields,[3,4] parameterized from experimental data, ab initio simulations, or machine learning techniques. However, in many cases, the transferability of such potentials can be poor, and classical MD fails to describe even qualitatively many interesting chemical and physical phenomena that are intrinsically nonclassical. The field of ab initio molecular dynamics (AIMD),[5–

[10] which is a quantum-classical approach, is between the two extremes of full quantum MD and classical MD. In addition to numerous studies of molecular properties, AIMD has been successfully applied to the modeling of materials, such as diffusion properties,[10] reaction processes,[12] vibrational frequencies,[13] and amorphous materials.[14] In AIMD, the nuclei are treated classically, but they are propagated on the potential energy surfaces (PESs) that are calculated "on-the-fly" quantum-mechanically through the solution of the time-independent Schrödinger equation (TISE) for the electrons. The interatomic forces are obtained as negative gradients of PESs. Compared to force fields, this approach allows for a better description of the time evolution of molecular systems as long as nuclear and electronic motions are not strongly coupled, and quantum effects between nuclei are not very strong.

The solution of TISE is central in the field of quantum chemistry. Unfortunately, it can be solved exactly only for very small systems due to the exponential growth of Hilbert space with molecular size. Many methods have been developed to find approximate solutions to TISE that can produce highly accurate results. However, they all require encoding quantum states on a classical computer, and without additional approximations, the computational cost of simulations scales exponentially. The classical full configuration interaction (FCI) method provides the exact solution for TISE for a given basis set but is not practical due to its exponential scaling with system size. More computationally efficient methods, such as density matrix renormalization group[15] or selected configuration interaction,[16] allow for larger simulations with FCI quality energies[17], but these methods still scale exponentially. The less accurate but computationally feasible method that has been used for the majority of AIMD simulations is density functional theory (DFT). Second-order perturbation theory (MP2) is also a common choice for AIMD. Another way to perform AIMD simulations is by using the quantum Monte Carlo method.[18,19]

In recent years, a new field of quantum computational chemistry has emerged with the idea of representing the state of a quantum chemical system on a quantum computer.[20–26] The idea to simulate quantum physics with quantum computers was theoretically proposed by Feynman.[27] Later, it was shown that the time evolution of quantum systems could be efficiently simulated on a fault-tolerant quantum computer.[28] Small scale demonstrations of such simulations have been performed on today's quantum computers.[29,30] When a fault-tolerant quantum computer with a large number of qubits is available, the quantum phase estimation (QPE) algorithm should be able to provide the exact ground state energies in polynomial time for systems too large to be simulated exactly on classical computers.[31,32] However, the number of qubits available on modern quantum computers is small, which limits the size of the system that can be simulated. Additionally, due to the low fidelity of the gates, especially 2-qubit CNOTs, the total number of gates in the circuits that can be run on quantum hardware is also limited. Thus, the application of QPE for problems intractable on classical computers is arguably years away.[33] For the noisy intermediate-scale quantum (NISQ)[33] hardware, one of the promising directions in quantum computational chemistry is the Variational Quantum Eigensolver (VQE) method.[34,35] It is based on the Rayleigh-Ritz variational principle that is widely used in quantum chemistry calculations on classical computers. VQE is one of the ways to avoid the problem of short coherence times of qubits in the NISQ quantum hardware devices by using classical variational optimization of the quantum circuit parameters.

In the present work, a quantum computer-based *ab initio* molecular dynamics method is used. In this method, the nuclei are propagated classically on PESs that are computed on-the-fly on a quantum computer using the VQE method. The forces for nuclear propagation are calculated numerically using the Hellmann-Feynman theorem (HFT) and a correlated sampling (CS) technique, a combination that allows one to compute numerical gradients with high precision and low cost. This method is extensively tested on IBM quantum hardware for the vibrational motion of the $H_2$ system. Moreover, the accuracy of the proposed gradient calculation approach is also estimated for larger molecules. Kassal *et al.*[36] have proposed an alternative approach to run quantum molecular dynamics on quantum computers. It directly applies kinetic and potential energy operators and bypasses the Born-Oppenheimer approximation. However, unlike VQE, it requires large-scale fault-tolerant quantum hardware. A similar AIMD study was performed by Sokolov *et al.*[37,38], a preprint for which was published within a few weeks after ours.

The paper is organized as follows. In Section 2, the used AIMD and VQE methodologies are described. In this section, we also describe the method we use for the calculation of accurate numerical gradients. In Section 3, the computational details are presented. Section 4 is devoted to the discussion of the results of AIMD dynamics for $H_2$ molecule and testing of the methods for larger molecules. In Section 5, the conclusions are presented.

## Methods
### Ab Initio Molecular Dynamics

In the present work, a standard quantum-classical approach within the Born-Oppenheimer approximation (BOA)[39] is employed to simulate AIMD. Classical trajectories (positions and momenta of the *N* nuclei) are propagated on an electronic PES according to the classical Hamilton's equations:

$$\dot{R}_I = \frac{P_I}{M_I} \tag{2a}$$

$$\dot{P}_I = F_I = -\nabla_I\big(E_0(R) + V_{NN}(R)\big) \tag{2b}$$

where $R_I$, $P_I$, $M_I$ are positions, momenta, and masses along the nuclear coordinate *I*, respectively, and *R* denotes all the nuclear coordinates. The force $F_I$ is computed as a negative gradient of the sum of the electronic ground state energy $E_0$ and nuclear repulsion energy $V_{NN}(R)$. Ground state electronic energy is computed "on-the-fly" at every step quantum mechanically through the solution of electronic TISE:

$$H_{el}\psi(r;R) = E\psi(r;R) \tag{3a}$$

$$H_{el} = T_e + V_{ee}(r) + V_{eN}(r,R) \tag{3b}$$

where $H_{el}$ is the electronic Hamiltonian, $T_e$ is the kinetic energy of electrons, $V_{ee}$ is the electron repulsion energy, and $V_{eN}$ is the potential energy of the electron-nuclear interaction. Each eigenvalue $E$ gives rise to a PES on which the nuclei can be propagated. In general, these PESs can be coupled through nonadiabatic effects.[11,40–43] In the present work, we consider only the ground state of the system, which corresponds to the smallest eigenvalue of the electronic Hamiltonian, and neglect nonadiabatic effects.

From the perspective of quantum chemistry, the computational aspect of AIMD simulations narrows down to the calculation of two quantities: gradients and energies. The former are used to carry out an MD trajectory by approximately solving Eq. (2), and the latter, along with the total kinetic energy, are used to evaluate the total energy as a check on the quality of the trajectory. The details of electronic structure calculations of energy and gradients are presented in the following sections. The classical propagation of the nuclei is straightforward and requires negligible computational resources compared to the electronic structure calculations. Since the dynamics simulations are performed "on-the-fly", we do not have access to full analytical PESs and Eq. (2) needs to be integrated numerically with a finite time step size. The choice of the integration method can significantly affect the number of energy and gradient calculations and accuracy. In this work, we use the Verlet algorithm in the velocity formulation.[44,45] This integration method requires only a single calculation of the forces per MD time step. Moreover, the Verlet algorithm has an extremely low drift of the total classical energy over long periods of time compared to other integrators, such as the Runge-Kutta integrator.[46]

## Ground state energy calculation

The exact solution of Eq. (3) is generally not available, and there are many quantum chemistry methods of varying accuracy and computational cost designed for classical computers to compute electronic energies and accurate analytical gradients. In this work, energies and gradients are computed using quantum computers. Due to the limitations of NISQ[33] hardware, the quantum computational chemistry field is, for the time being, mostly based on using the VQE method and its variations. Calculation of the ground state energy $E_0$ for a system described by electronic Hamiltonian $H_{el}$ is performed using the variational principle that is widely used in quantum chemistry. When applied in VQE, this method optimizes the quantum circuit parameters that are used for quantum state preparation on a quantum computer:

$$E_0 \leq \frac{\langle \psi(\theta)|H_{el}|\psi(\theta)\rangle}{\langle \psi(\theta)|\psi(\theta)\rangle} \quad (4)$$

where $\psi(\theta)$ is the trial wave function, and the quantity on the right of Eq. (4) is minimized iteratively with respect to the parameter(s) $\theta$. The trial state $\psi(\theta)$ is obtained from the initial state $\psi$ by applying a unitary operator $\psi(\theta) = U(\theta)\psi_0$.

To find the ground state energy of the system $E_0$, the electronic Hamiltonian is represented in the second quantized form,[47]

$$H_{el} = \sum_{p,q} h_{pq} a_p^\dagger a_q + \sum_{p,q,r,s} h_{pqrs} a_p^\dagger a_q^\dagger a_r a_s \quad (5)$$

where $a_q^\dagger$ and $a_q$ are creation and annihilation operators for the various spin orbitals $q$. The $h_{pq}$ are one-electron integrals representing the kinetic energy of the electrons and their Coulomb interaction with the nuclei, and $h_{pqrs}$ are two-electron integrals representing electron-electron Coulomb repulsion. One-electron and two-electron integrals can be efficiently computed on a classical computer in any quantum chemistry package. Next, the variational form of the wave function, an ansatz, needs to be chosen. The design of ansatzes for VQE calculation is an active area of research, varying from the so-called hardware-efficient[34,48] (HE) ansatzes that are designed to be simpler and easier to run on NISQ hardware to chemically inspired ones such as the unitary coupled cluster (UCC) ansatz.[35] An adaptive ansatz[49] can significantly reduce the computational cost by the addition and removal of operators in the set. The UCC method was initially applied to solve problems in physics[50–52], and with the development of NISQ quantum hardware, it was proposed for solving quantum computational chemistry problems.[22,34] In the UCC method, the initial state is obtained from a reference Hartree-Fock state by applying the exponential unitary operator:

$$|\psi_{UCC}\rangle = e^{\hat{T}-\hat{T}^{\dagger}}|\psi_{HF}\rangle \qquad (6)$$

where $T$ is the excitation operator that excites electrons from the occupied orbitals $\alpha$ in the reference configuration into virtual (unoccupied) orbitals $i$. In the truncated version (UCCSD), which is used in present work, only single and double excitations are considered:

$$\hat{T} = \hat{T}_1(\theta) + \hat{T}_2(\theta) = \sum_{i \in virt, \alpha \in occ} \theta_{i\alpha}\, a_i^{\dagger} a_{\alpha} + \sum_{i,j \in virt, \alpha, \beta \in occ} \theta_{ij\alpha\beta}\, a_i^{\dagger} a_j^{\dagger} a_{\alpha} a_{\beta} \qquad (7)$$

The UCC method is intractable on classical computers due to exponential scaling, but it can be solved in polynomial time on a quantum computer.[35] UCC is a modification of the classical coupled cluster (CC) method, which is considered a gold standard in quantum chemistry[53] and can be efficiently solved on a classical computer. In the standard CC method, the wave function is obtained by using the non-unitary $e^{\hat{T}}$ operator $|\psi_{CC}\rangle = e^{\hat{T}}|\psi_{HF}\rangle$. The UCC method retains the advantages of the CC method and has an additional beneficial property of being variational, unlike the standard CC method. In addition to the standard UCCSD there are modified versions of the method, such as generalized UCCGSD.[54]

To simulate a chemical system in the second quantized representation, the creation and annihilation operators acting on indistinguishable fermions need to be mapped into operators acting on distinguishable qubits. The most common encoding methods are Jordan-Wigner,[55] parity,[56] and Bravyi-Kitaev[57] encoding, all of which produce Hamiltonians of the following form:

$$H = \sum_j \alpha_j P_j = \sum_j \alpha_j \prod_i \sigma_i^j \qquad (8)$$

where $P_j$ are Pauli strings, $\sigma_i^j$ are Pauli matrices, and $\alpha_j$ are coefficients that depend on the values of electron integrals $h_{pq}$ and $h_{pqrs}$. Each encoding type has its own advantages. For example, the parity mapping takes advantage of the system symmetry and reduces the number of qubits required for the simulation. Jordan-Wigner encoding makes it easier to enforce the conservation of the number of particles in a simulation. The Bravyi-Kitaev mapping yields qubit Hamiltonians with a logarithmically smaller number of Pauli terms compared to Jordan-Wigner and parity encodings.

### Gradient calculation

Analytical gradients can be computed through the solution of coupled-perturbed Hartree-Fock equations on a classical computer and using the wave function optimized on a quantum computer to obtain the gradient matrix elements.[58,59] In the present work, the gradients are computed numerically using central differences. In the most straightforward brute force approach, for each of the $3N$ nuclear degrees of freedom $x$, the energies at displaced geometries $x - d$ and $x + d$ are computed using full VQE optimization. The first and most obvious disadvantage of this approach is that two additional VQE optimizations have to be performed for each coordinate per one time step of MD simulation. More importantly, the stochastic nature of quantum measurements has a strong negative effect on the accuracy of numerical gradients. Evaluating energies at displaced geometries $\langle\psi(x+d)|H(x+d)|\psi(x+d)\rangle$ and $\langle\psi(x-d)|H(x-d)|\psi(x-d)\rangle$ on a quantum device involves two measurements. The resulting two quantities have errors due to the stochastic nature of the quantum measurements, and these errors are independent. The small displacement $d$ in the denominator typically has to be a number on the order of $10^{-3}$ $a_0$ to $10^{-4}$ $a_0$, which results in the amplification of gradient error by orders of magnitude. Thus, the brute force approach to compute gradients numerically for MD simulations on a quantum computer is not feasible due to the astronomical numbers of measurements required to obtain accurate gradients.

When the HFT is applied, the force can be computed using the following expression:

$$F = -\frac{d}{dx}\langle\psi(x)|H|\psi(x)\rangle \approx -\left\langle\psi(x)\left|\frac{dH}{dx}\right|\psi(x)\right\rangle$$
$$\approx -\frac{\langle\psi(x)|H(x+d)|\psi(x)\rangle - \langle\psi(x)|H(x-d)|\psi(x)\rangle}{2d}. \quad (9)$$

This approach allows optimization of the wave function only once per MD time step. A naïve inspection of eq 9 suggests that one needs to perform two quantum measurements to compute the gradient in this approach. Therefore, it suffers from the same precision problem as the brute force approach because errors at the displaced geometries are not correlated. However, upon closer inspection of matrix elements in eq. (9), the Pauli strings in Eq. (8) stay the same at displaced geometries, only coefficients $\alpha$ change. This allows us to employ a third approach with HFT and correlated sampling (CS), as shown in Fig. 1. In this scheme, the HFT is utilized by using the optimized circuit parameters at point $x$, $\theta(x)$, to evaluate energy at $x - d$ and $x + d$. In addition, instead of performing independent measurements, the results of sampling the Pauli strings of qubit-mapped Hamiltonian $\sum_j \alpha_j^x P_j^x$ (see Eq. (8)) at the central point $x$ are used for energy evaluation at $x - d$ and $x + d$. In other words, the same set of shots is used to evaluate the expectation values of

$H(x + d)$ and $H(x - d)$. Thus, the only extra calculations at displaced geometries are the computations of electron integrals $h_{pq}$ and $h_{pqrs}$ on a classical computer. Apart from the obvious advantage of performing a smaller number of computations (only one full VQE cycle) on a quantum computer, this scheme has a feature that is more important for numerical gradient calculation. The energies at displaced geometries $\langle\psi(x)|\sum_j \alpha_j^{x-d} P_j^x|\psi(x)\rangle$ and $\langle\psi(x)|\sum_j \alpha_j^{x+d} P_j^x|\psi(x)\rangle$ are obtained from the quantum measurements at the same point. As a result, a much smaller error is expected from division by a small number in Eq. (9).

It is important to note the errors associated with HFT gradients. First, if $\psi(x)$ is an exact eigenfunction of $H$, the HFT is an exact expression for the gradient (upper Eq. (9)). For typical quantum chemistry applications, as pointed out by Pulay,[60,61] when $\psi(x)$ is not an exact eigenstate, there is a contribution to the gradient that is neglected in the HFT expression owing to the quantum chemistry bases being atom-centered and thus changing with $x$. This omitted term can be large, particularly for single determinant (e.g., Hartree-Fock) approximations. Bakken et al.[62] carried out a careful analysis and showed how increasing the basis set level and degree of configuration interaction can lead to significant improvements in HFT-based gradients. In our application to $H_2$, with minimal basis but with full configuration interaction, we have verified that the HFT errors (including those associated with the finite difference approximation) are insignificant regarding the generation of trajectories that conserve energy to a good degree. The difference between equilibrium distance of $H_2$ computed using exact diagonalization and the UCCSD with HFT is around $10^{-6}$ Å for STO-3G basis (0.734862 Å and 0.734863 Å respectively). In the era beyond NISQ devices, when hardware capabilities allow deeper circuits and longer coherence times, analytical gradients, which do not have the discussed error, could be used in MD simulations.[58,63]

However, in our simulations, these errors are negligible compared to the stochastic errors associated with limited samples on quantum computers. Until we go beyond NISQ devices in terms of fidelity and the number of qubits, the numerical estimation of gradients is the only option. We believe the gain in precision of numerical gradients by using HFT with CS is much higher than the error associated with the neglect of Pulay forces.

Although the number of evaluations of numerical gradients grows with the system size as $6N$, where $N$ is the number of nuclei, our gradient computation procedure has a very low computational overhead compared to just the energy evaluation. There are no additional expensive VQE optimizations, and only a calculation of electron integrals for each displaced geometry is required, the cost of which is significantly lower than additional energy evaluations.

Since the $H_2$ molecule contains only 2 electrons, single and double excitations in UCCSD ansatz cover all possible electronic excitations. Thus, the UCCSD method is equivalent to FCI for the studied system and, therefore, provides the exact solution of the TISE for a given basis set. We start by running UCCSD simulations using the state vector simulator Aer in Qiskit package,[64] which calculates the full electronic wave function and does not involve measurements. Such simulations are good for benchmarking because there is no noise present, either stochastic, from the probabilistic nature of quantum measurements or noise from the quantum hardware due to decoherence effects.

It has been shown by Wecker *et al.*[65] that the number of quantum measurements $N_{shots}$ required to estimate the energy to precision $\varepsilon$ has an upper bound of

$$N_{shots} = \frac{\left(\sum_i |\alpha_j|\right)^2}{\varepsilon^2} \qquad (10)$$

where $\alpha_j$ are the Pauli string coefficients in the Hamiltonian, which depend on the electron integrals.

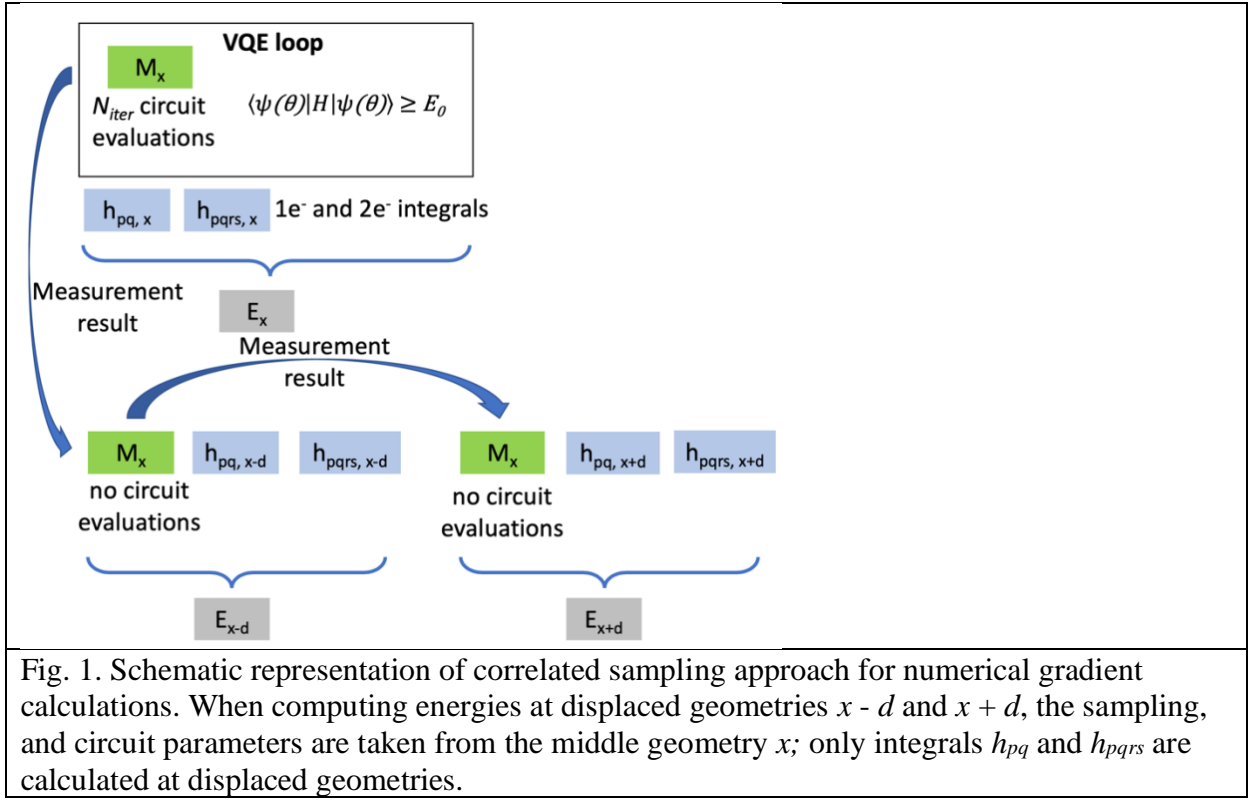

Fig. 1. Schematic representation of correlated sampling approach for numerical gradient calculations. When computing energies at displaced geometries $x - d$ and $x + d$, the sampling, and circuit parameters are taken from the middle geometry $x$; only integrals $h_{pq}$ and $h_{pqrs}$ are calculated at displaced geometries.

There are multiple sources of error involved in the gradient calculation. First, VQE optimization might converge to a local minimum instead of the global one, which leads to variational parameters that overestimate the energy. Second, the chosen ansatz might not be able to recover all-electron correlation energy and, therefore, be unable to recover the exact energy for a given basis. To eliminate all errors not related to the gradient calculation procedure and to assess the accuracy of the utilized gradient calculation, we use the following strategy. First, direct diagonalization of the Hamiltonian is performed to obtain the exact wave function. Then fermionic operators are mapped to qubit operators, and the resulting Hamiltonian is of the form as in Eq. (8). Consequently, Pauli strings of the mapped Hamiltonian are used to compute the matrix elements $\langle \psi(x) \rangle$ and $\langle \psi(x) \rangle$. Then, all matrix elements are sampled. Finally, all sampled matrix elements are summed to obtain total energies at $x - d$ and $x + d$ to compute the gradient at $x$.

## Computational Details

In this study, all VQE calculations were performed using Qiskit 0.19.6 package.[64] This package is developed by IBM and used to run calculations on IBM quantum devices or simulators. IBM quantum devices are provided by the Quantum Computing User Program located in Oak Ridge Leadership Computing Facility (OLCF).[66] We used a slightly modified version of Qiskit, which adds the implementation of the CS technique. The optimization of variational parameters $\theta$ in VQE was done using the COBYLA optimizer. Calculations on hardware were done on IBM 5-qubit quantum device Vigo. One- and two-electron integrals were computed using PySCF[67] electronic structure program, which is interfaced with Qiskit. All calculations were performed using the STO-3G Gaussian basis set. The Hamiltonian was mapped to qubits using parity mapping.[68] It was previously shown[69,70] how the number of qubits required for describing a fermionic system could be reduced by using symmetry, effectively reducing the size of Hilbert space. For example, the $H_2$ molecule in a minimal basis set has been studied before with the number of qubits reduced to two from the standard four qubits. The resulting circuit for measurement in the computational basis is shown in Figure 2a.

We use an approach developed by Bravyi *et al.* to further reduce the number of qubits needed to run calculations on a quantum computer down to only one qubit.[56] The one-qubit circuit is depicted in Figure 2b. This approach of "tapering off" qubits is based on $Z_2$ symmetries. The idea is to find Pauli strings that commute with the Hamiltonian. Such strings represent the symmetries of the Hamiltonian. Using these symmetries one can construct a unitary operator that transforms the Hamiltonian so that the transformed Hamiltonian acts trivially or with $\sigma_x$ on a set of qubits. Thus, such a set of qubits can be "tapered off" (left out of the simulation). The process described above can be thought of as a projection of the Hamiltonian into symmetry subspaces, which can be simulated with fewer qubits. For the $H_2$ molecule the simplification is achieved by using Brillouin theorem, according to which the single-excitation amplitudes are zero. Additionally, $D_{\infty h}$ point group symmetry allows one to further reduce the required number of qubits. The method of "tapering off" qubits was generalized further improved by Setia et al.[71]

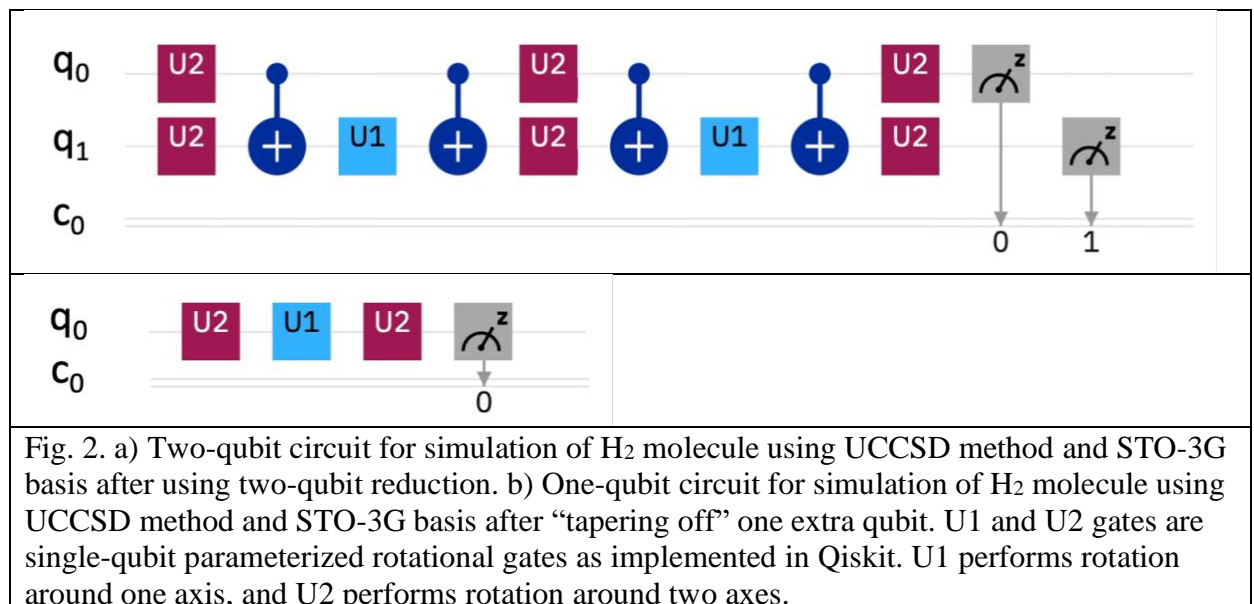

Fig. 2. a) Two-qubit circuit for simulation of $H_2$ molecule using UCCSD method and STO-3G basis after using two-qubit reduction. b) One-qubit circuit for simulation of $H_2$ molecule using UCCSD method and STO-3G basis after "tapering off" one extra qubit. U1 and U2 gates are single-qubit parameterized rotational gates as implemented in Qiskit. U1 performs rotation around one axis, and U2 performs rotation around two axes.

## Results and Discussion
### Accuracy of energies and gradients

Before running the circuits on the real quantum hardware, we start with the calculations using the QASM simulator.[72] This simulator mimics the measurement on actual quantum hardware by using a random number generator to obtain the result of measurement (0 or 1) for each qubit. QASM calculations allow us to estimate the actual stochastic error for the $H_2$ molecule. We ran VQE simulations with the number of shots ranging from $8 \times 10^3$ to $512 \times 10^3$. The energy errors were calculated as the difference between the VQE energy and the exact energy obtained from direct Hamiltonian diagonalization. It can be seen from Fig. 3 that at least $256 \times 10^3$ shots are needed to keep the error in energy within the chemical accuracy range of 1.6 mHa (1 kcal/mol).

For a smaller number of measurements, chemical accuracy cannot be guaranteed. The errors are expected to increase on real quantum hardware due to the device noise.

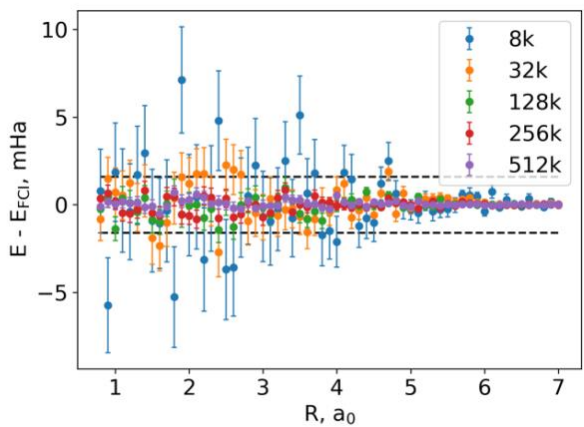

Fig. 3. Errors in the potential energy of $H_2$ molecule computed on a QASM simulator for the 1-qubit UCCSD/STO-3G circuit. Colors represent the number of shots per Pauli string, from $8 \times 10^3$ to $512 \times 10^3$. Error bars correspond to one standard deviation. The black horizontal line represents the error threshold for chemical accuracy $\varepsilon = 1.6 \times 10^{-3}$ Ha (1 kcal/mol).

Implementation of the numerical gradients requires a careful choice of the displacement parameter $d$ in Eq. (9) because division by a number much smaller than 1 significantly amplifies the errors in energy. The errors in gradients $G = \frac{E(R+d)-E(R-d)}{2d}$ computed with different $d$ values are presented in Fig. 4. In Fig. 4, the black line corresponds to the gradients calculated using energies obtained from the state vector simulator, where the wave function was optimized at displaced geometries (by $d$), i. e. without application of HFT. The lines related to the state vector calculation using HFT for different $d$ values are indistinguishable from the black line. The insert in Fig. 4 shows the errors in gradients for state vector simulations. Smaller $d$ values produce more accurate gradients, but for all $d$ values in this range, the errors are on the order of $10^{-6}$ $Ha/a_0$, which is three orders of magnitude smaller than the errors introduced by stochastic noise in calculations using QASM simulator with $512 \times 10^3$ shots (Fig. S1a in Supplementary Information). Fig. S1a shows that stochastic noise is three orders of magnitude larger than the noise coming from *d* even when CS is used. As expected, the error decreases with an increasing number of shots consistent with $O(\frac{1}{\sqrt{N}})$ estimate using Eq. (10) (see Fig. S1b).

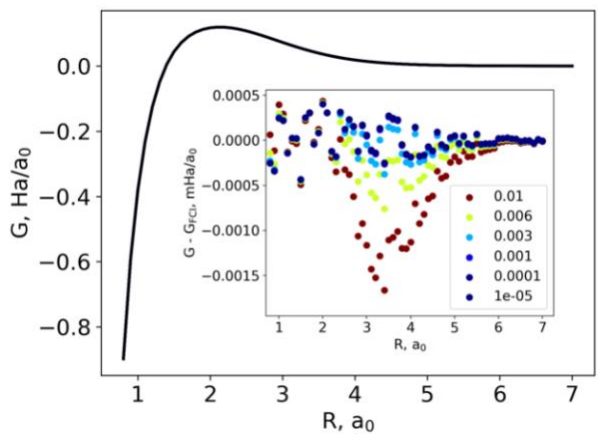

Fig. 4. Numerical gradient G for 2 qubit UCCSD circuit computed with state vector simulator for different values of displacement *d* after application of the HFT. The black line corresponds to the calculation without the HFT application, which was used to calculate absolute errors of the gradient (colored dots in the insert). Colors in the insert represent different *d* values.

## Simulations of larger molecules

The scalability of our approach for calculating gradients numerically for larger molecular systems was tested by performing gradient calculations using FCI with STO-3G basis set for the following molecules: $H_2$, $H_4$, $LiH$, $BeH^+$. The number of qubits increases from the 1-qubit $H_2$ system up to the 12-qubit $BeH^+$ molecule. Systems with a different number of qubits for a given molecule correspond to simulations with and without symmetry application to reduce the number of qubits. In Fig. 5, we compare the number of shots per Pauli string per VQE iteration required to reach precision in gradients of $\varepsilon = 1$mHa/$a_0$ relative to the FCI solution. To obtain the number of shots required to achieve a certain accuracy, the errors were calculated for the number of shots per Pauli string ranging from 64 to $6.5\times10^7$, fitted using the least squares method according to the $\varepsilon = \frac{\alpha}{\sqrt{N_{shots}}}$ formula. After that, the number of shots was extrapolated to achieve the required accuracy

(see Fig. S4 for details). It is clear that the only viable option to compute numerical gradients is to use HFT with CS because the number of shots required to reach the same precision without applying HFT is higher by 4 orders of magnitude across all studied molecules. Higher counts for BeH$^+$ molecule can be explained by the larger coefficients $\alpha_j$ in front of Pauli strings (equation 10) compared to the rest of the molecules. The errors potentially introduced by neglecting Pulay forces are negligible compared to stochastic uncertainties when HFT is not applied. The smallest number of shots required without CS is on the order of $10^7$. This is a very large number of shots to execute on NISQ hardware. The scaling of such calculations makes it unpractical even for a single-point energy calculation, and it is cost-prohibitive for AIMD simulations on NISQ-era hardware. To make it worse, the number of shots is multiplied by the number of Pauli strings in the Hamiltonian, which grows with system size, and by the number of iterations in the VQE optimization. In Fig. 5 the plotting of shots per Pauli string and not the total number of shots is intentional to separate the increase in shots due to larger gradient errors from the increase in shots due to the number of Hamiltonian terms. In addition, the number of measurements for a certain number of Hamiltonian terms can vary because commuting terms can be measured simultaneously. There are algorithms to reduce the number of measurements through a grouping of Pauli strings[73], but using them is beyond the scope of this work.

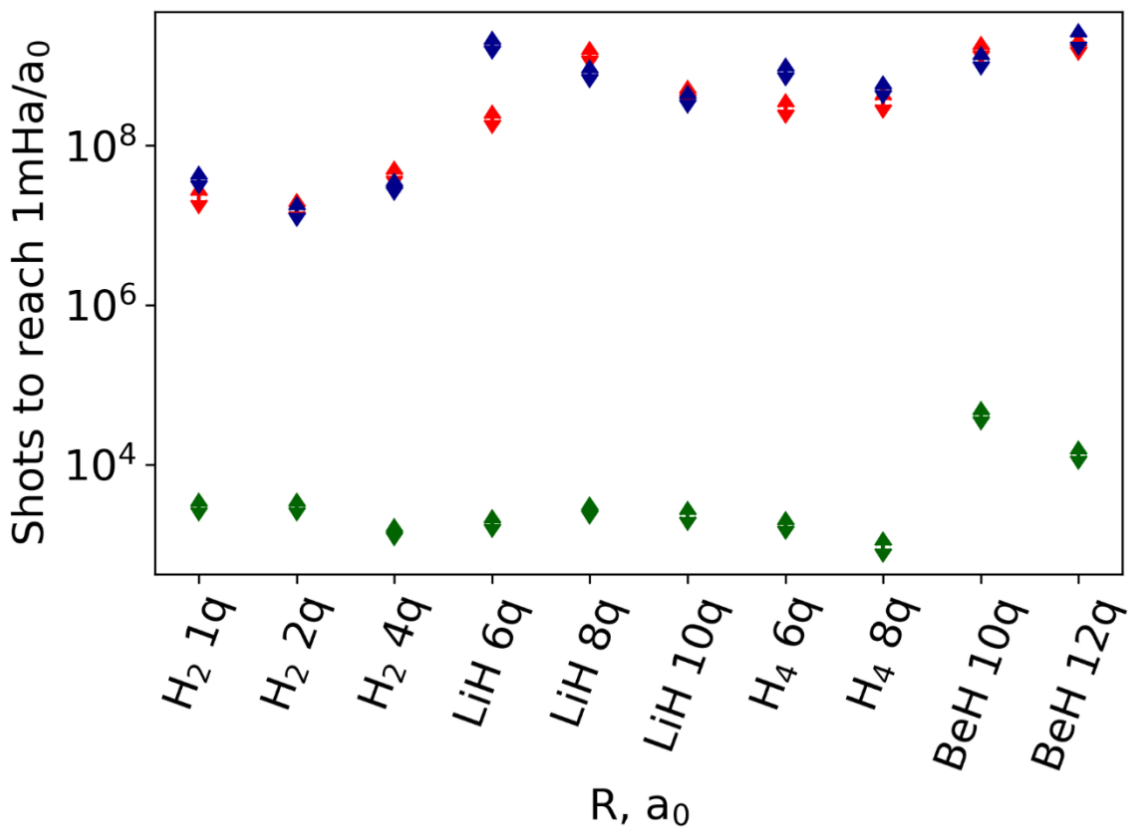

Fig. 5. The number of shots per Pauli string required to achieve accuracy $\varepsilon = 1\text{mHa}/a_0$ for different molecules computed at $R = 4.0\ a_0$. Green color corresponds to calculations with HFT and CS. Blue represents calculations with applied HFT but without CS, while red corresponds to calculations without HFT and CS. Arrows represent error bars.

It is important to point out that we use VQE with UCCSD ansatz as one of the options for AIMD on quantum computers. The method we use for computing gradients will work with other methods that measure Hamiltonian directly. It is known that classical optimization in the VQE method can become an intractable problem when the number of parameters is large. The number of variational parameters grows fast when using standard UCCSD ansatz. However, various techniques can be used to reduce the number of parameters through the elimination of parameters that correspond to excitations not contributing to correlation energy. For example, one could use an adaptive approach like ADAPT-VQE,[49,74] which drastically reduces the number of variational parameters. Alternatively, one could use a recently developed quantum solver of contracted eigenvalue equations which improves upon the variational quantum eigensolver in terms of circuit depth and built-in energy gradients.[75]

## $H_2$ simulations on IBM Q devices

The quality of an MD trajectory for a 1D system can be represented by plotting positions vs. momenta (Fig. 6). When a molecule returns to its starting position, the trajectory is represented by the full ellipse and covers the whole configuration space. If the initial and final points coincide with each other, the total energy is conserved, and therefore accurate simulation of longer trajectories is possible. When the total energy decreases over time due to accumulated errors, the trajectory drifts from the ellipse towards the center, and two atoms collapse on each other. If the total energy increases, the ellipse spirals out, and the molecule eventually dissociates. Thus, the goal of an accurate MD scheme is to minimize the total energy drift over time.

For simulations on hardware, IBM Q Vigo was chosen due to the low readout and single-qubit errors. A single MD trajectory for $H_2$ molecule computed on the IBM Q Vigo machine with $295 \times 10^3$ shots and measurement error mitigation, as implemented in the Qiskit Ignis[64] package, is shown in Fig. 6. Fig. 6a presents the trajectory corresponding to a complete vibration of $H_2$ in phase space. The trajectory calculated on a classical computer using the state vector simulator is represented by orange lines. The green line corresponds to simulations without VQE optimization where the variational parameters are preoptimized using the state vector simulator, essentially reducing the use of a quantum computer for sampling quantum circuits with optimal parameters. The blue lines correspond to the trajectory obtained using full VQE optimization at every step. The starting geometry for MD trajectory is $x = 2$ $a_0$ and initial velocity is $v_0 = 0$. Time step of 5 a. u. was used for numerical integration of equations of motion. The deviation of the total energy of the trajectory obtained using the state vector simulator from the ideal trajectory is minimal, with a slight increase of 0.03 mHa in the region of short internuclear distances, which happens around t = 5 fs. However, when the molecule leaves a short internuclear distance region, the total energy goes back to the initial value. This error is attributed to the finite time step numerical integration, and the accuracy can be systematically improved by decreasing the time step. Due to the noise (stochastic and decoherence from the quantum device) in trajectories calculated on the quantum device (green and blue lines), the total energy value oscillates around zero with the amplitude around 6 mHa (see Fig. 6b); however, there is no drift during the simulation. It is consistent with Figure 6a) where the internuclear distance difference between the beginning and the end of the full VQE trajectory (dashed blue line) is 0.01 $a_0$. The potential energy and gradient values obtained from hardware simulations oscillate around the FCI values and reproduce the exact results well (see Fig. S2). It is important to point out that the relative error of circuit parameter $\theta$ along the trajectory is much larger than the errors in energies and gradients, which can be seen in Fig. 5b. This can be explained by the fact that the parameters optimized using the noiseless state vector simulator might not correspond to the minimal energy of the system with noise on an actual quantum device. Also, the gradient of energy with respect to parameter $\theta$ is small, i.e., the deviation of $\theta$ from optimal values does not translate into energy deviations comparable with hardware noise error.

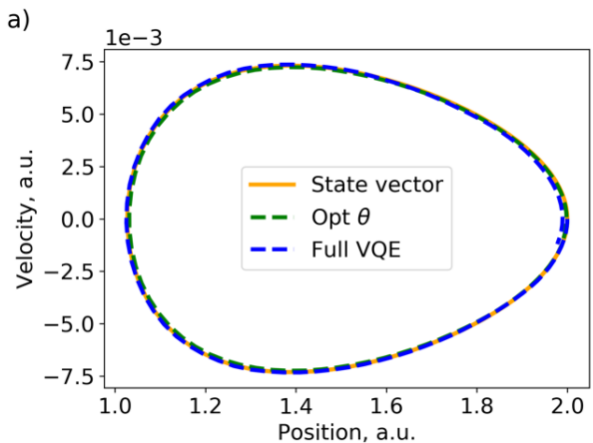

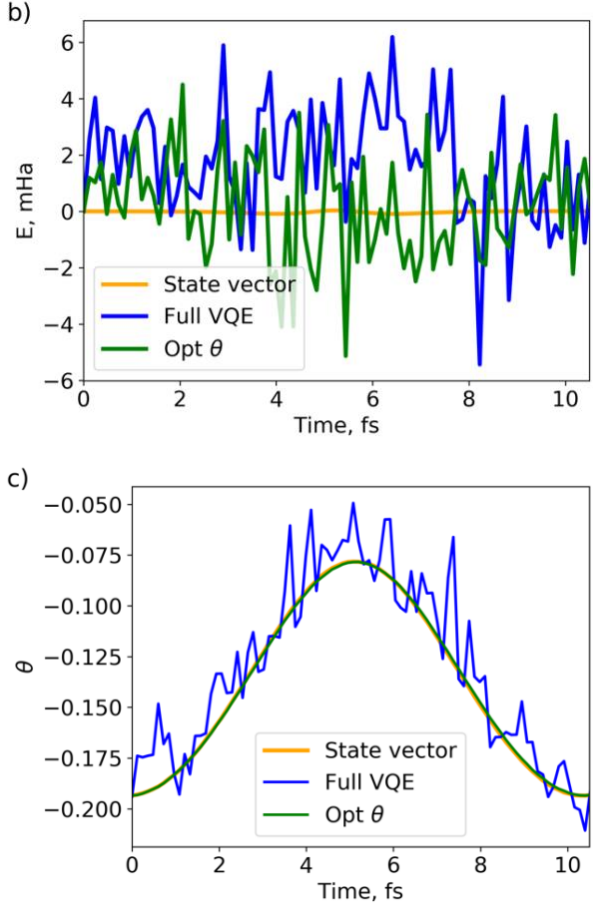

Fig. 6. A single loop MD trajectory for a 1-qubit $H_2$ system computed using UCCSD ansatz. The orange line corresponds to the state vector simulation: a) position vs. velocity, b) total energy vs. time, c) parameter $\theta$ vs. time. Blue lines represent the simulation on the IBM Q Vigo machine using $295 \times 10^3$ shots with measurement error mitigation. Green lines correspond to simulations on the IBM Q Vigo machine where parameters were optimized using a state vector simulator, and sampling was done using $295 \times 10^3$ shots with measurement error mitigation.

In addition to the 1-qubit $H_2$ system, we performed simulations on a 2-qubit system with 1 parameter $\theta$ on the IBM Q Vigo machine. The 2-qubit system does not produce more accurate results compared to a 1-qubit system, even if the hardware error rate was much lower, because the symmetry reduction used to obtain a 1-qubit system does not affect accuracy. However, we provide results for a 2-qubit system to estimate the results associated with MD simulations for less trivial circuits. From Fig. 7b, it can be seen that the conservation of energy for 2-qubit systems is significantly worse than for the 1-qubit system. However, in the 2-qubit system trajectory, the potential energy is significantly overestimated, and the deviation from FCI energy is relatively constant over all internuclear distances. This potential energy shift in the 2-qubit system is easily explained by the hardware noise from four CNOT gates, which are not present in the 1-qubit system; each of the CNOT gates has an error rate on the order of 1%. This noise can be mitigated using various techniques, for example, zero-noise extrapolation,[76–78] symmetry verification[79], or reduced density matrix purification.[80] However, such simulations on the quantum computers available today for the whole MD trajectory would be very time-consuming.

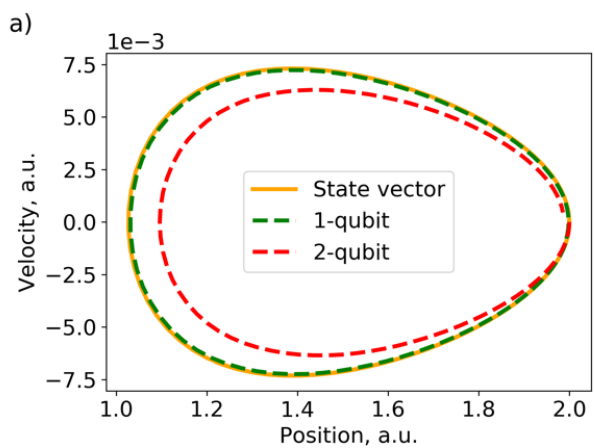

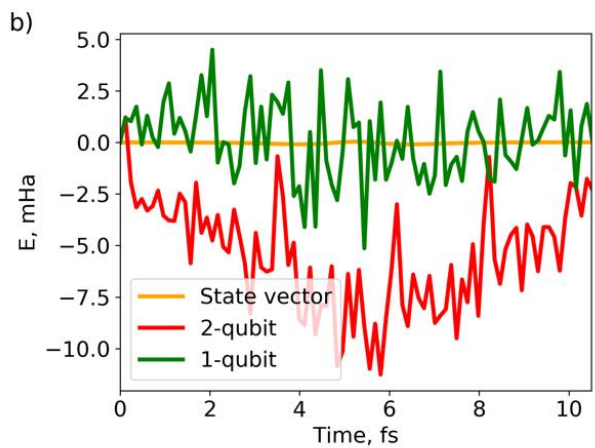

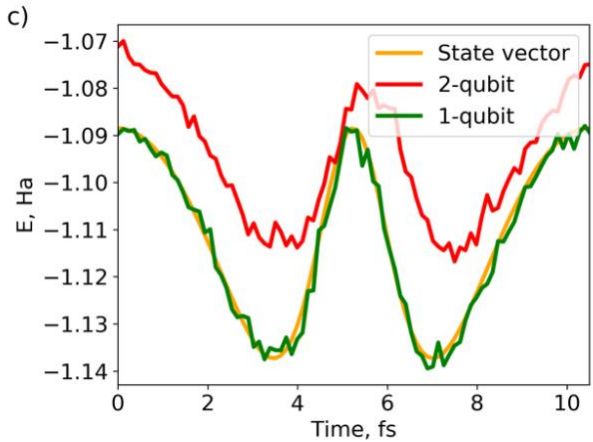

Fig. 7. A single loop MD trajectory for a $H_2$ system computed using UCCSD ansatz on IBMQ Vigo machine using $295 \times 10^3$ shots with measurement error mitigation: a) position vs. velocity, b) total energy vs. time, c) potential energy vs. time. The orange line corresponds to the state vector simulation (solutions for 2-qubit and 1-qubit systems match). Red and green lines represent the hardware simulations of the 2-qubit and 1-qubit systems, respectively. Parameters used in simulations on hardware were optimized with state vector simulation, VQE optimization was not performed.

## Conclusions

In this work, we illustrated an approach to AIMD simulations on NISQ-era quantum hardware, which relies on the VQE method for the calculation of ground-state electronic energies. Compared to the energy calculations, AIMD simulations on quantum computers pose an extra challenge and need accurate gradient calculations compared to classical computers. Algorithms for efficient analytical gradient calculations fully on a quantum computer will require quantum hardware beyond NISQ-era. In this work, we performed AIMD simulations using numerical gradients computed with finite differences. Numerical gradients computations on a quantum computer without extra approximations require a cost-prohibitive number of measurements to achieve the precision required for AIMD simulations. This is due to the stochastic errors associated with independent energy measurements on a quantum computer and the linearly increasing number of full energy evaluations with system size. To avoid the cost-prohibitive scaling, we used the Hellman-Feynman theorem for the energy gradient calculations in this work, which is a further approximation. In addition, we employed a straightforward correlated sampling (CS) technique, which takes advantage of the fact that qubit Hamiltonians at displaced geometries are expanded using the same Pauli strings as the central point where energy is computed. The CS technique allows one to perform only one quantum measurement at a central point and only recompute electron integrals at displaced geometries.

As a proof of concept and to establish how practical AIMD simulations are on modern IBM quantum hardware, we chose to model the $H_2$ system using 1-qubit and 2-qubit circuits. Even though there is no total energy drift overall and trajectories return to the initial positions, the total classical energy fluctuates by several mHa over the course of simulations. This is still beyond the chemical accuracy threshold, although we should note that a qualitatively correct trajectory resulted. Therefore, quantum hardware with lower error rates in the future is needed to improve the quality of energies and gradients, and therefore provide more accurate trajectories. As expected, the simulations based on 2-qubit circuits provide trajectories of inferior quality and show that additional error mitigation is required to simulate AIMD of larger systems on modern NISQ quantum hardware.

Although our example of vibrational motion in $H_2$ is elementary, it demonstrates that AIMD simulations on NISQ quantum hardware are possible. However, significant hardware improvements are needed to perform simulations on larger molecular systems. Until fault-tolerant quantum computers are available that would enable analytical gradient calculations, AIMD on quantum computers has to be run using numerical gradients. Until then, potential improvements can be made in designing more efficient variational ansatzes to reduce the cost and increase the precision of AIMD simulations on quantum hardware.

## Supplementary Material

The supplementary material contains the additional figures referenced in the article.

## Acknowledgments

This material is based upon work supported by the U.S. Department of Energy, Office of Science, Office of Fusion Energy Sciences, under Award Number DE-SC0020249. This research used resources of the Oak Ridge Leadership Computing Facility at the Oak Ridge National Laboratory, which is supported by the Office of Science of the U.S. Department of Energy under Contract No. DE-AC05-00OR22725. This work was performed, in part, at the Center for Nanoscale Materials, a U.S. Department of Energy Office of Science User Facility, and supported by the U.S. Department of Energy, Office of Science, under Contract No. DE-AC02-06CH11357. The authors would like to thank Mark Kostuk, Stefan Bringuier, and Pejman Jouzdani from General Atomics, as well as Ahren W. Jasper from Chemical Sciences and Engineering Division and Martin Suchara from Mathematics and Computer Science Division at Argonne National Laboratory for fruitful discussions.

## Data Availability

Data that support the findings of this study are available upon request.

## Author Contributions

D.A.F., M.J.O., S.K.G. and Y.A. designed the research. D.A.F. wrote the code, performed simulations on simulator and hardware and analyzed the results. All authors wrote the manuscript.

## Code Availability

Modified version of Qiskit package is available upon request.